\def\lesssim{\mathrel{\hbox{\rlap{\hbox{\lower4pt\hbox{$\sim$}}}\hbox{$<$}}}}

\def\gtrsim{\mathrel{\hbox{\rlap{\hbox{\lower4pt\hbox{$\sim$}}}\hbox{$>$}}}}

\def\lteff{log ${T_{\rm eff}}$~}

\def\ll_lsun{log$({L/\rm L_{\odot}})$~}

\def\masa_msun{$M/ \rm M_{\odot}$~}

\def\m_mstar{$M/M_{*}$~}

\def\rxj{RX\,J2117.1+3412}
\newcommand{\Teff}{$T\mathrm{\hspace*{-0.4ex}_{eff}}$}
\newcommand{\logg}{$\log\,g$\hspace*{0.5ex}}

\documentclass{aa}
\usepackage{graphicx}
        
\begin{document}

\title{Asteroseismological constraints on the pulsating planetary 
nebula nucleus (PG1159-type) RX\,J2117.1+3412}

\author{A. H. C\'orsico$^{1,2}$\thanks{Member of the Carrera del Investigador
Cient\'{\i}fico y Tecnol\'ogico, CONICET, Argentina.},
L. G. Althaus$^{1,2\star}$,
M. M. Miller Bertolami$^{1,2,3}$\thanks{Fellow of CONICET, Argentina.} 
\and K. Werner$^4$}

\offprints{A. H. C\'orsico}

\institute{
$^1$   Facultad   de   Ciencias  Astron\'omicas   y   Geof\'{\i}sicas,
Universidad  Nacional de  La Plata,  Paseo del  Bosque S/N,  (1900) La
Plata, Argentina.\\ $^2$ Instituto de Astrof\'{\i}sica La Plata, IALP,
CONICET-UNLP\\  $^3$ Max-Planck-Institut f\"ur  Astrophysik, Garching,
Germany\\  $^4$ Institut f\"ur Astronomie und Astrophysik, Universit\"at
T\"ubingen, Sand 1, 72076 T\"ubingen, Germany\\ 
\email{acorsico,althaus,mmiller@fcaglp.unlp.edu.ar; 
werner@astro.uni-tuebingen.de} }
  
\date{Received; accepted}

\abstract{}{We present asteroseismological inferences on 
\rxj, the  hottest known pulsating PG1159 star.
Our  results are  based on  full PG1159  evolutionary  models recently
presented  by  Miller Bertolami  \&  Althaus  (2006).}  {We  performed
extensive  computations  of adiabatic  $g$-mode  pulsation periods  on
PG1159 evolutionary models with stellar masses ranging from $0.530$ to
$0.741  M_{\odot}$.   PG1159 stellar  models  are  extracted from  the
complete evolution of progenitor  stars started from the ZAMS, through
the thermally pulsing  AGB and born-again phases to  the domain of the
PG1159 stars.  We  constrained the stellar mass of  \rxj\ by comparing
the  observed period spacing  with the  asymptotic period  spacing and
with the average of the computed period spacings. We also employed the
individual  observed periods  to find  a  representative seismological
model  for   \rxj.}{We  derive  a  stellar   mass  $M_*\sim  0.56-0.57
M_{\odot}$ from the period spacing data alone. In addition, we found a
best-fit model representative for
\rxj\ with an effective temperature $T_{\rm eff}= 163\,400$ K, 
a stellar mass $M_*= 0.565  M_{\odot}$, and a surface gravity $\log g=
6.61$.   The   derived  stellar   luminosity  and  radius   are  $\log
(L_*/L_{\odot})= 3.36$ and $\log(R_*/R_{\odot})= -1.23$, respectively,
and the  He-rich envelope thickness is $M_{\rm  env}= 0.02 M_{\odot}$.
We derive a seismic distance $d \sim  452$ pc and a linear size of the
planetary nebula $D_{\rm PN} \sim  1.72$ pc.  These inferences seem to
solve the discrepancy between the \rxj\ evolutionary timescale and the
size of the  nebula.  All of the seismological tools  we use concur to
the  conclusion that  \rxj\ must  have a  stellar mass  $M_*\sim 0.565
M_{\odot}$, much in agreement with recent asteroseismology studies and
in  clear   conflict  with   the  predictions  of   spectroscopy  plus
evolutionary tracks.}{}
 
\keywords{stars:  evolution ---  stars: interiors  --- stars: oscillations 
--- stars: variables: other (GW Virginis)--- white dwarfs}

\authorrunning{C\'orsico et al.}

\titlerunning{Asteroseismological constraints on \rxj}

\maketitle

 
\section{Introduction}

Pulsating   PG1159   stars   (or   GW   Vir  stars)   are   very   hot
hydrogen-deficient  post-Asymptotic  Giant  Branch  (AGB)  stars  with
surface layers  rich in  helium, carbon and  oxygen (Werner  \& Herwig
2006).  They exhibit  multiperiodic luminosity variations with periods
in the  range $300-3000$ seconds, attributable  to nonradial pulsation
$g$-modes.  Some GW  Vir stars are still  embedded in a planetary
nebula (commonly called PNNVs; Planetary Nebula Nucleus Variable), and are
characterized  by much  higher luminosity  than the  ``naked''  GW Vir
stars  (those without  nebulae). PG1159  stars are  thought to  be the
evolutionary   link   between  post-AGB   stars   and   most  of   the
hydrogen-deficient white  dwarfs. These stars  are believed to  be the
result of a born again episode  triggered either by a very late helium
thermal  pulse (VLTP)  occurring in  a hot  white dwarf  shortly after
hydrogen burning  has almost ceased (see  Sch\"onberner
1979 and more recently Herwig  2001, Lawlor \& MacDonald 2003, Althaus
et  al.  2005,  and Miller  Bertolami et  al. 2006)  or a  late helium
thermal  pulse (LTP) that  takes place  during the  post-AGB evolution
when  hydrogen  burning  is  still  active  (see  Bl\"ocker  2001  for
references).   During  a  VLTP  episode,  most  of  the  hydrogen-rich
envelope of  the star is  burnt in the helium  flash-driven convection
zone, while in a LTP  the hydrogen deficiency is the result of
a dilution episode.   In both cases, the star  returns rapidly back to
the AGB and finally into the domain of high effective temperature as a
hydrogen-deficient, quiescent helium-burning object. This helium 
burning phase continues until the hot white-dwarf domain is reached.

\rxj\ is  the hottest  known  pulsating PG1159  star. It  was
discovered as an  X-ray source in  the  ROSAT sky survey  (Motch et al.
1993). The star is embedded  in an old, extended and diffuse planetary
nebula,   which at the distance of 1400 pc (Appleton et al. 1993) 
makes it the  largest  presently  known  with a  diameter  of  5.3  pc.
This implies an expansion age much in excess as compared with that 
predicted by stellar evolution (Appleton et  al. 1993). For the central  
star Werner et  al. (1996) and Rauch \& Werner (1997) 
derived,  with the help of optical, UV, and X-ray spectra and 
NLTE model atmospheres, $T_{\rm  eff}= 170 \pm 10$  
kK and  $\log g=  6^{+0.3}_{-0.2}$, and a surface composition  
of He/C/O= 47.5/23.8/6.2 (by number of
atoms). Furthermore,  a  strong  mass-loss rate of  $\dot{M}  \approx
10^{-7}-10^{-8}   M_{\odot}$/yr  has   been  measured   (Koesterke  et
al. 1998; Koesterke
\& Werner 1998). The central star of \rxj\ was
discovered  to  be  a  pulsator  independently by  Watson  (1992)  and
Vauclair  et al.  (1993), with pulsation periods that are intermediate 
between those of PNNVs and  naked GW Vir stars.   
The effective  temperature of  the central star  defines the blue  
edge of  the observational  PG1159 instability
strip.   Vauclair  et al.   (2002)  (hereinafter  VEA02) published  an
amazing  richness of  asteroseismological results  on  \rxj\
from  a  multisite  photometric  campaign that included  two  Whole 
Earth Telescope (WET, Nather et al. 1990) runs.
Without  actually performing pulsation  computations, they  were still
able to derive  fundamental quantities for \rxj,  such as the
stellar mass,  the helium-rich envelope mass  fraction, rotation rate,
luminosity, and distance.

It is  important to note  that at the  time the analysis of  VEA02 was
performed,  no  realistic  PG1159  evolutionary  models  suitable  for
interpreting  high-luminosity,  high-temperature transition objects
such  as  \rxj\  were  available. In  fact,  while  the  $0.7
M_{\odot}$ evolutionary  sequence of Wood \& Faulkner  (1986) available 
at that time fits the
location of  \rxj\ in the  $\log g - \log  T_{\rm eff}$ plane,
its predicted  surface abundance composition is  not representative of
the observed abundances in \rxj.  More importantly, the Wood
\& Faulkner (1986) models do not constitute appropriate structures for
PG1159 models that have been  evolved through a born-again episode.
On the  other  hand, the  evolutionary
tracks of Gautschy (1997), with surface abundances more appropriate to
PG1159 stars,  do not have  luminosities high enough to  be compatible
with \rxj.  In view of the lack of realistic stellar models
for a full asteroseismological analysis,  VEA02 were forced to use the
scaling relation between period spacing and mass of Kawaler
\& Bradley (1994) to infer the stellar mass of \rxj.  It is
worth  noting  that  this  relation  was  derived  for  GW  Vir  stars
characterized by much lower luminosities than those of the PNNVs.

One of the results of VEA02 is that the asteroseismological mass 
they derive for \rxj\ is about $0.56
M_{\odot}$, in serious disagreement with the mass derived from 
spectroscopical analysis ($0.72 M_{\odot}$). This fact has motivated 
us to undertake the 
present investigation taking full advantage of the new generation of 
PG1159 evolutionary models recently developed by Miller  
Bertolami \& Althaus (2006).  In fact, these authors have computed realistic
PG1159   evolutionary  sequences  for   different  stellar   masses  by
considering the  complete evolution  of  progenitor  stars. They
 have followed  in detail all of the  evolutionary phases prior
to the  formation of PG1159 stars, particularly  the born-again stage.
We believe that the  evolutionary models presented by Miller Bertolami
\& Althaus (2006) represent  a solid theoretical background to analyze
the evolutionary  and pulsational status  of hot PG1159 stars  like
\rxj, a transition object  for which stellar models  calculated with
artificial evolutionary procedures should be taken with caution.

In  this  work  we  perform  an  asteroseismological  analysis  of 
\rxj\ based on the PG1159 evolutionary models of Miller Bertolami
\&  Althaus   (2006).  Emphasis  is  placed   on  deriving  pertinent 
structural parameters for  this star by  using its observed period  
spectrum.  In particular,  we derive  the stellar  mass for  this object  
from three
different seismological approaches. Additionally, the implications 
of our results for the evolutionary status of \rxj\ and for its
planetary nebula are discussed. The  paper is organized as follows:
in  the next  Section we  briefly describe  the input  physics  of the
PG1159 evolutionary models and the  pulsational code we employ.  In \S
\ref{period-spacing} we  derive the stellar  mass of \rxj\ by
means  of the observed  period spacing.    In Section  \ref{fitting} we
derive structural  parameters of this  star by employing  the individual 
observed periods. In \S \ref{masses} we discuss the discrepancy 
we found between the seismological and the spectroscopic 
mass of \rxj, and in \S \ref{helium} we discuss the 
thickness of the outer envelope of the star as predicted by our 
seismological model. The asteroseismological distance to \rxj\ 
and the implications of our analysis for 
its planetary nebula are provided in Section \ref{pn}.  
Finally,  in Sect.  \ref{conclusions} we summarize
our main results and make some concluding remarks. 

\section{Evolutionary models and numerical tools}
\label{evolutionary}

The pulsation analysis presented in this work relies on stellar models
that  take into account  the complete  evolution of  PG1159 progenitor
stars. The pulsation stability  properties of these PG1159 models have 
recently been analyzed by  C\'orsico et al. (2006).  The  stages for 
the formation and evolution of PG1159 stars were  computed with  the  
LPCODE  evolutionary  code, which  is described in detail  in Althaus 
et al.  (2005).   Briefly, LPCODE uses
OPAL radiative opacities  (including carbon- and oxygen-rich mixtures)
from the compilation of Iglesias \& Rogers (1996), complemented at the
low-temperature regime  with the  molecular opacities of  Alexander \&
Ferguson  (1994)  (with  solar  metallicity).   Chemical  changes are 
performed  via  a  time-dependent  scheme that  simultaneously  treats
nuclear evolution and mixing  processes due to convection, salt finger
and   overshooting.   Convective   overshooting  is  treated   as  an
exponentially   decaying  diffusive  process   above  and   below  any
convective region.

\begin{figure}
\centering
\includegraphics[clip,width=250pt]{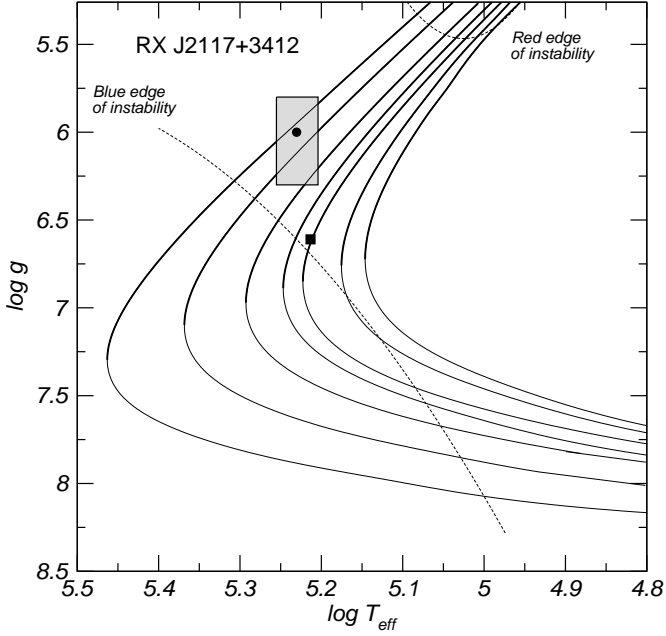}
\caption{The Miller Bertolami \& Althaus (2006) 
PG1159 full evolutionary  tracks in  the $\log T_{\rm  eff} -\log  g$ plane,
corresponding  to stellar  masses  of (\emph{from  right  to left}):  $0.530,
0.542, 0.565,  0.589, 0.609$,  $0.664$, and $0.741  M_{\odot}$.  Thick
(thin)  curves correspond to  stages before  (after) the  models reach
their highest effective temperature.   The black circle with the error
box  is  the  location  of  \rxj\  according  to  spectroscopy
($T_{\rm eff}= 170 \pm 10$  kK and $\log g= 6^{+0.3}_{-0.2}$) and
the  black square is the  location of  the star  
as predicted  by our asteroseismological  analysis (\S \ref{fitting}).  
The  blue (hot) and red (cool) boundaries  of  the theoretical  dipole
($\ell= 1$) instability domain  according to C\'orsico et al.  (2006) 
are depicted with thin dashed lines.}
\label{teff-g}
\end{figure}

Specifically, the background of stellar models were extracted from the
PG1159 evolutionary calculations recently presented in Miller Bertolami \&
Althaus (2006) (0.53, 0.542, 0.565, 0.609 and 0.664 $M_{\odot}$
sequences), Althaus et al. (2005) (0.589 $M_{\odot}$) and C\'orsico et
al. (2006) (0.741 $M_{\odot}$). The range of initial masses of these
models on the ZAMS is $1-3.75 M_{\odot}$. We refer the reader 
to those works for
details. All of the sequences were followed from the ZAMS 
 (assuming solar metallicities) through the
thermally pulsing and mass-loss phases on the AGB.  After experiencing
several thermal pulses, the progenitors departed from the AGB and
evolved towards high effective temperatures.  The mass-loss rate
during the AGB was arbitrarily kept fixed, as to obtain a final helium
shell flash during the early white dwarf cooling phase (VLTP).  As a
result, evolution proceeds through the born-again stage, which brings
the remnants back to the giant domain. After this episode, most of the
hydrogen content of the stars was burnt, and the hydrogen-deficient
remnants evolved at constant luminosity to the domain of PG1159 stars
with surface chemical composition rich in helium, carbon and
oxygen. The evolutionary tracks in the $\log T_{\rm eff} - \log g$
plane are displayed in Fig. \ref{teff-g}. For the 1-$M_{\odot}$ sequence 
two different AGB evolutions
were considered, with different mass-loss rates as to obtain different
numbers of thermal pulses and, eventually, two different remnant
masses of 0.530 and $0.542 M_{\odot}$.  Mass-loss episodes after the
VLTP were not considered in the PG1159 evolutionary sequences we
employed here, despite the fact that mass-loss episodes for 
PG1159 stars have been reported (see Werner \& Herwig 2006). 

We computed adiabatic pulsation periods with the help of the numerical
code we employed in our  previous works (see C\'orsico \& Althaus 2006
for details).   We analyzed about  3000 PG1159 models covering  a wide
range of effective temperatures ($5.4 \gtrsim
\log(T_{\rm eff}) \gtrsim 4.8$) and luminosities ($0 \lesssim
\log(L_*/L_{\odot}) \lesssim 4.2$), and a range of stellar masses
($0.530 \leq  M_*/M_{\odot} \leq 0.741$).  For each  model we computed
$\ell= 1$  $g$-mode periods\footnote{Quadrupole ($\ell= 2$) modes 
have not been detected in \rxj\ (VEA02).} in  the range 
$50 \lesssim  \Pi_k \lesssim
3600$ s, which  comfortably covers the period spectrum  observed in 
\rxj\ ($690 - 1200$ s).

\section{Mass determination from the observed period spacing}
\label{period-spacing}

Here, we constrain the stellar  mass of \rxj\ by comparing the
asymptotic  period  spacing  ($\Delta  \Pi_{\ell}^{\rm  a}$)  and  the
average of the  computed period spacings ($\overline{\Delta \Pi_{k}}$)
with   the   \emph{observed}   period   spacing   ($\Delta   \Pi^{\rm
O}$)\footnote{Note that the vast majority of asteroseismic studies are
based on the asymptotic period spacing to infer the stellar mass of GW
Vir pulsators.}.  These methods take  full advantage of the  fact that
the  period  spacing of  PG1159  pulsators  depends  primarily on  the
stellar mass (Kawaler \& Bradley 1994; C\'orsico \& Althaus 2006).

\begin{figure}
\centering
\includegraphics[clip,width=250pt]{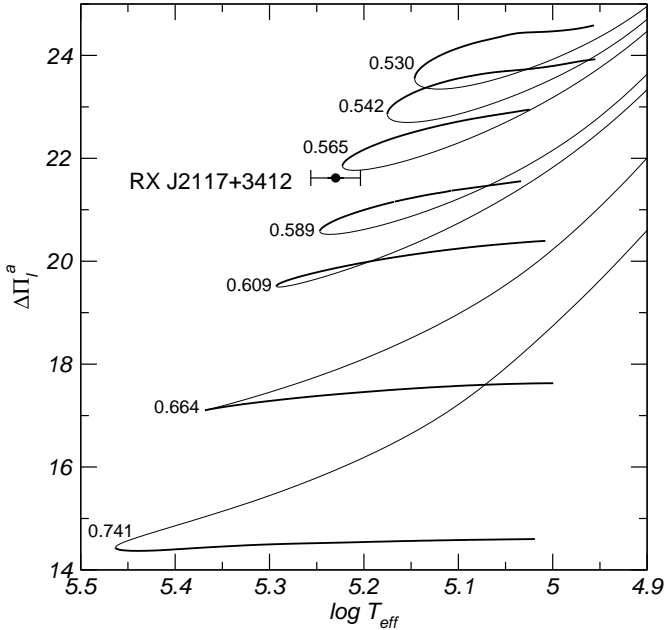}
\caption{The dipole ($\ell= 1$) asymptotic period spacing 
($\Delta \Pi_{\ell}^{\rm a}$) for different stellar masses in terms of
the effective  temperature.  Numbers at the left of each curve denote the 
stellar masses (in solar units). Thick (thin) curves  correspond to stages
before (after)  the models reach their  highest effective temperature.
The small black  circle with the error bar depicts  the location of 
\rxj\ ($T_{\rm eff}= 170 \pm 10$ kK and $\Delta \Pi^{\rm
O}= 21.618 \pm 0.008$ s).}
\label{aps}
\end{figure}

\begin{figure}
\centering
\includegraphics[clip,width=250pt]{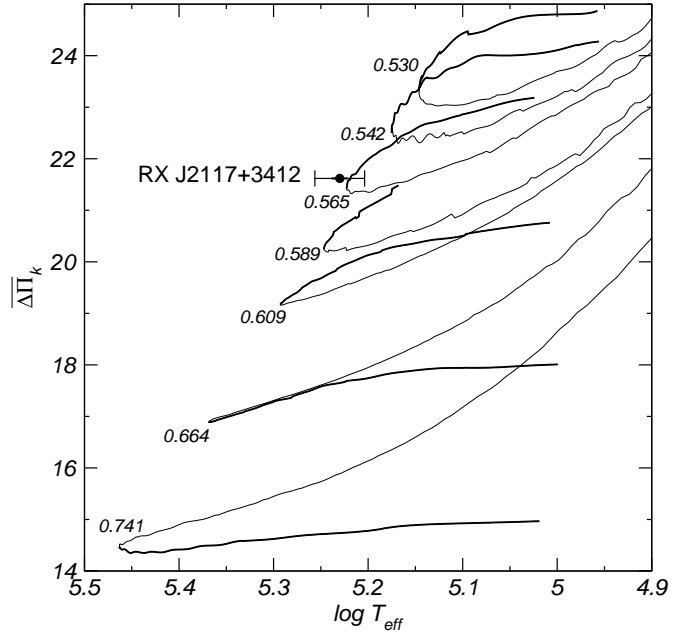}
\caption{Same as Fig. \ref{aps}, but for the average 
of the computed period spacings ($\overline{\Delta \Pi_{k}}$).}
\label{psp}
\end{figure}

The asymptotic period  spacing and the average of  the computed period
spacings  for  $\ell=  1$  modes   as  a  function  of  the  effective
temperature   are  displayed  in   Figs.   \ref{aps}   and  \ref{psp},
respectively,  for  different  stellar  masses. Also  shown  in  these
diagrams is the location of \rxj, with $T_{\rm eff}= 170
\pm 10$ kK (Werner et al.  1996) and $\Delta \Pi^{\rm O}= 21.618 \pm
0.008$ s (VEA02).  Here, $\Delta \Pi_{\ell}^{\rm a}=
\Pi_0 / \sqrt{\ell(\ell+1)}$, where $\Pi_0= 2 \pi^2 [ \int_{r_1}^{r_2}
(N/r) dr]^{-1}$, being  $N$ the Brunt-V\"ais\"al\"a frequency (Tassoul
et al. 1990).  The  quantity $\overline{\Delta \Pi_{k}}$, on the other
hand, is  assessed by averaging  the computed forward  period spacings
($\Delta \Pi_{k}=  \Pi_{k+1}- \Pi_{k}$) in  the range of  the observed
periods in \rxj.

From   a  comparison   between  $\Delta   \Pi^{\rm  O}$   and  $\Delta
\Pi_{\ell}^{\rm a}$  (``first method''), we  obtain a stellar  mass of
$M_*= 0.568_{-0.003}^{+0.008} M_{\odot}$.  The quoted uncertainties in
the  value  of  $M_*$  come  from  the  errors  in  the  spectroscopic
determination  of the  effective  temperature\footnote{Our uncertainties 
are different from those quoted in VEA02 which are due entirely to the 
fact that they lack evolutionary calculations appropriate for \rxj.}.  
Note  that, since  the
$M_*=  0.565   M_{\odot}$  evolutionary  track  does   not  reach  the
spectroscopic $T_{\rm  eff}$ of \rxj\ (see Figs. \ref{teff-g}
and  \ref{aps}), we  are forced  to extrapolate  the value  of $\Delta
\Pi_{\ell}^{\rm a}$ for that effective temperature ($170$ kK) and
also for its upper limit ($180$ kK) to derive the stellar mass.

In the same way, we get $M_*= 0.560_{-0.013}^{+0.018} M_{\odot}$ if we
compare $\Delta \Pi^{\rm O}$ and $\overline{\Delta \Pi_{k}}$ (``second
method'').    Again,   we   need   to  extrapolate   the   values   of
$\overline{\Delta  \Pi_{k}}$ for  the spectroscopic  value  of $T_{\rm
eff}$ and its upper limit in  order to derive the stellar mass and its
lower limit.

It is evident that both estimations of the stellar mass nearly agree
\footnote{Note that in both derivations of $M_*$ we have 
assumed  that  \rxj\  is evolving  towards  higher  effective
temperatures, before it reaches the evolutionary knee
in the  $\log T_{\rm eff}-  \log g$ diagram.  In other words,  we have
used  the  thick  portions  of  the curves  in  Figs.   \ref{aps}  and
\ref{psp} to infer the stellar mass. Assuming instead that
the star is on the beginning of its white dwarf cooling phase, then 
the thin portions of the  curves should have been considered. In that case
we would have  obtained values of the stellar  mass only slightly different
from those quoted  in the text.}.  The slightly  higher value of $M_*$
as derived from $\Delta
\Pi_{\ell}^{\rm a}$ is due to that usually the asymptotic period
spacing  is a  bit  larger than  the  average of  the computed  period
spacings (see C\'orsico  \& Althaus 2006).  The first method to derive 
the stellar mass is  somewhat less realistic than the second one, 
because  the asymptotic predictions are, in principle,
only valid for chemically homogeneous stellar models, while our PG1159
models are indeed chemically stratified (see \S \ref{helium}).

VEA02 derive a stellar  mass of $M_*= 0.56_ {-0.04}^{+0.02} M_{\odot}$
by  extrapolating  $\Pi_0$ from  the  formula  of  Kawaler \&  Bradley
(1994).  It  is interesting to  note that, despite  that extrapolation
being based on a relation which is valid for luminosities between $1.6
\leq \log(L_*/L_{\odot}) \leq 2.8$  (much lower than the luminosity of
\rxj),  the mass  derived by VEA02  is in  complete agreement
with our predictions, although our uncertainties are much smaller.

Finally, we note that our  inferred stellar mass value of $M_* \approx
0.56-0.57  M_{\odot}$, is  in  serious conflict  with  the value  $M_*
= 0.72_{-0.12}^{+0.15} M_{\odot}$ as derived  from spectroscopy coupled  to 
evolutionary tracks\footnote{The quoted uncertainties in the 
spectroscopic value of $M_*$ have their origin in  
the uncertainties in the determination of 
$\log g$ and $T_{\rm eff}$, e.g.,  we neglect possible 
uncertainties in the evolutionary computations.} 
(Miller Bertolami \& Althaus 2006). This will be  discussed in more 
detail below.

\section{Constraints from the individual observed periods}
\label{fitting}

\begin{figure}
\centering
\includegraphics[clip,width=250pt]{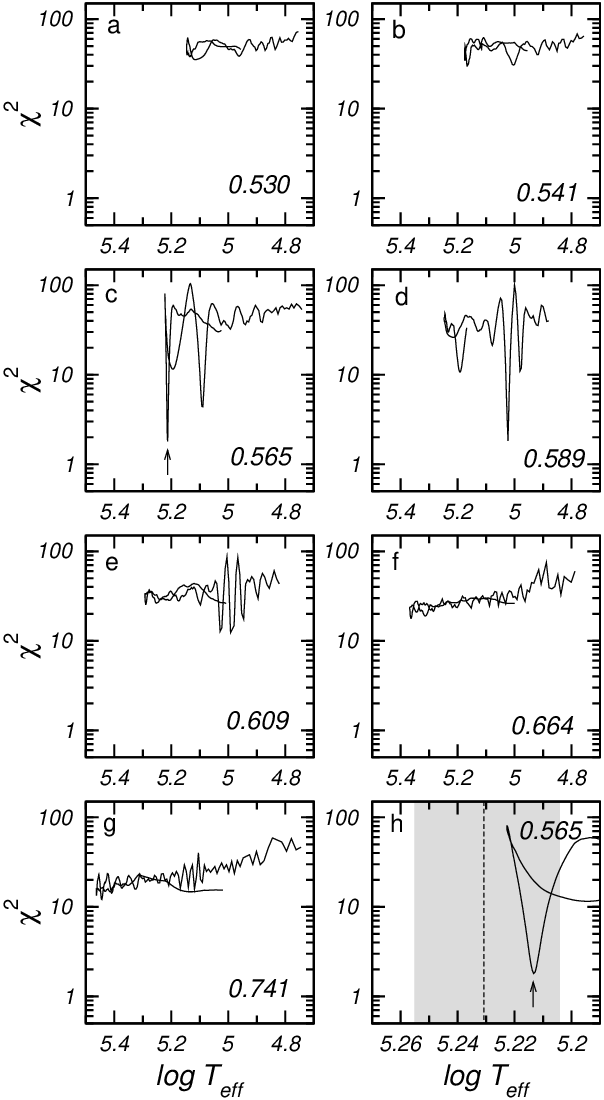}
\caption{The quality function of the period fit, $\chi^2$, 
in  terms of  $T_{\rm eff}$  for the  PG1159 sequences  with different
stellar masses,  indicated (in solar mass) at  the right-bottom corner
of each panel.  Note the strong  minima in panels {\bf c} and {\bf d},
corresponding to  $M_*= 0.565  M_{\odot}$ and $M_*=  0.589 M_{\odot}$,
respectively. Panel  {\bf h} is a  zoom of the region  with the strong
minimum seen in panel {\bf c} (arrow).  The vertical dashed line is the
spectroscopic $T_{\rm eff}$ of \rxj\ (170 kK) and 
the grey zone depicts its uncertainties ($\pm 10$ kK).}
\label{chi2} 
\end{figure}

In  this approach  we seek  a pulsation  model that  best  matches the
individual pulsation  periods of \rxj.  The goodness  of the
match between  the theoretical  $\ell= 1$ pulsation  periods ($\Pi_k$)
and the observed  periods ($\Pi_i^{\rm O}$) is measured  by means of a
quality function defined as $\chi^2(M_*, T_{\rm eff})=
\sum_{i=1}^{n} \min[(\Pi_i^{\rm O}- \Pi_k)^2]/n$, where 
$n$ (= 20) is the number of observed periods (first column in Table
\ref{tabla1}).  The PG 1159 model that shows the lowest value of
$\chi^2$ will be adopted as the ``best-fit'' model.

We evaluate the function $\chi^2(M_*, T_{\rm eff})$ for stellar masses
$0.530, 0.542,  0.565, 0.589, 0.609, 0.664$,  and $0.741 M_{\odot}$,
whereas for  the effective temperature  we employed a much  more finer
grid ($\Delta T_{\rm eff}= 10-30$  K).  The function $\chi^2$ in terms
of the effective temperature for  different stellar masses is shown in
the mosaic of  Fig. \ref{chi2}. We found only  two important minima of
comparable  magnitude, one  of them  for  the model  with $M_*=  0.565
M_{\odot}$ and  $T_{\rm eff}\approx 163.4$ kK (panel {\bf  c}) and
the other  one for the model  with $M_*= 0.589  M_{\odot}$ and $T_{\rm
eff} \approx 105$  kK (panel {\bf d}). Another  minimum, albeit less
pronounced,  is  also  encountered  for  the model  with  $M_*=  0.565
M_{\odot}$ at  $T_{\rm eff}\approx 123$ kK. 
We adopt the  
first model as the best-fit model because
it has an effective temperature  relatively close to that suggested by
spectroscopy.  A detailed comparison of the observed $m= 0$ periods of
\rxj\  with the theoretical  periods of the best-fit  model is
given in  Table \ref{tabla1}.  The high  quality of our  period fit is
quantitatively  reflected  by  the  average  of  the  absolute  period
differences, $\overline{\delta \Pi_i}= (\sum_{i=1}^n |\delta
\Pi_i|)/n$, where $\delta \Pi_i= \Pi_i^{\rm O} -\Pi_k$, and by the
root-mean-square  residual,   $\sigma_{_{\delta  \Pi_i}}=  \sqrt{(\sum
|\delta \Pi_i|^2)/n}  $.  We obtain $\overline{\delta  \Pi_i}= 1.08$ s
and $\sigma_{_{\delta \Pi_i}}= 1.34$ s.  The quality of our fit for 
\rxj\ is  substantially better than those achieved  by Kawaler \&
Bradley (1994) and C\'orsico \& Althaus (2006) ($\overline{\delta
\Pi_i}=  1.19$  s  and
$\overline{\delta \Pi_i}= 1.79$ s, respectively) for PG 1159-035.

\begin{table}
\centering
\caption{Observed $m= 0$ periods ($\Pi_i^{\rm O}$) of \rxj\ 
(taken  from  VEA02), theoretical  $\ell=  1$  periods ($\Pi_k$),  
period   differences  ($\delta   \Pi_i$),  radial orders ($k$),  
normalized  growth   rates ($\eta_k$), and rates of 
period change ($\dot{\Pi}_k$) of the best-fit model.}
\begin{tabular}{cccccc}
\hline
\hline
$\Pi_i^{\rm  O}$  &  $\Pi_k$ &  $\delta  \Pi_i$  &  $k$ &  $\eta_k$  &
$\dot{\Pi}_k$\\ 
$[$s$]$ & $[$s$]$ & $[$s$]$ & & $[10^{-7}]$ & $[10^{-11}$ s/s$]$\\
\hline
$692.27$ & $689.77$ & $2.5$  & 30 & $-0.25$ & $-2.99$ \\  
$712.98$ & $711.61$ & $1.37$ & 31 & $-0.23$ & $-5.38$ \\ 
$733.95$ & $732.36$ & $1.59$ & 32 & $-0.15$ & $-5.52$ \\ 
$757.35$ & $754.47$ & $2.88$ & 33 & $ 0.18$ & $-5.85$ \\ 
$778.92$ & $776.99$ & $1.93$ & 34 & $ 0.10$ & $-5.02$ \\ 
$799.50$ & $798.55$ & $0.95$ & 35 & $ 0.62$ & $-3.29$ \\
$821.15$ & $820.45$ & $0.7$  & 36 & $ 0.78$ & $-3.22$ \\  
$843.69$ & $843.05$ & $0.64$ & 37 & $ 1.75$ & $-6.00$ \\ 
---      & $865.34$ & ---    & 38 & $ 3.15$ & $-8.14$ \\ 
$885.74$ & $886.67$ & $-0.93$ & 39 & $ 3.68$& $-7.76$ \\ 
$907.49$ & $907.18$ &  $0.31$ & 40 & $ 5.54$ & $-8.47$ \\
--- & $929.26$ & --- & 41 & $ 6.75$ & $-6.26$ \\ 
$951.75$ & $952.70$ & $-0.95$ & 42 & $ 8.44$ & $-4.84$ \\ 
$972.25$ & $974.32$ & $-2.07$ & 43 & $ 12.8$ & $-8.73$ \\ 
$994.39$ &  $994.46$ & $-0.07$ & 44 & $ 16.5$ & $-11.76$ \\ 
$1016.47$ & $1016.43$ & $0.04$ & 45 & $ 20.6$ & $-10.94$ \\ 
$1038.12$ & $1038.84$ & $-0.72$ & 46 & $ 24.8$ & $-11.12$ \\
$1058.03$ &  $1059.87$ & $-1.84$  & 47  & $ 23.3$  & $-8.72$ \\ 
 --- & $1081.23$ & ---  & 48 & $  25.5$ & $-8.25$ \\ 
$1103.29$  & $1103.12$ & $0.17$ & 49 & $ 31.9$ &  $-14.24$ \\ 
$1124.12$ & $1124.68$ & $-0.56$ & 50 & $  27.8$ & $-13.66$ \\ 
$1146.35$  & $1146.53$ & $-0.18$ &  51 & $ 19.2$ & $-10.07$ \\ 
--- & $1167.32$ & --- & 52 & $ 11.2$ & $-12.32$ \\
$1189.96$ & $1188.85$ & $1.11$ & 53 & $-2.27$ & $-10.27$ \\
\hline
\hline
\end{tabular}
\label{tabla1}
\end{table}

Table \ref{tabla1}  also shows the  linear growth rates ($\eta_k$) of  
the fitted
pulsation  modes  (fifth  column),  computed  with the  
nonadiabatic pulsation  code described in C\'orsico et  al. (2006). 
A positive value of $\eta_k$ implies pulsational instability.
It is interesting to note that the domain of unstable-mode periods 
($757 \lesssim \Pi_k \lesssim  1167 $ s) of our best-fit model
nearly  coincides  with  the  range  of the  observed  periods  for  
\rxj, although our calculations  predict the observed modes with
$k= 30, 31, 32$ and  $53$ to be pulsationally stables, and conversely,
modes such as  $k= 38, 41, 48$ and $52$,  which are unstable according
to our predictions, are not observed at all in \rxj.

\begin{figure*}
\centering
\includegraphics[width=0.9\textwidth]{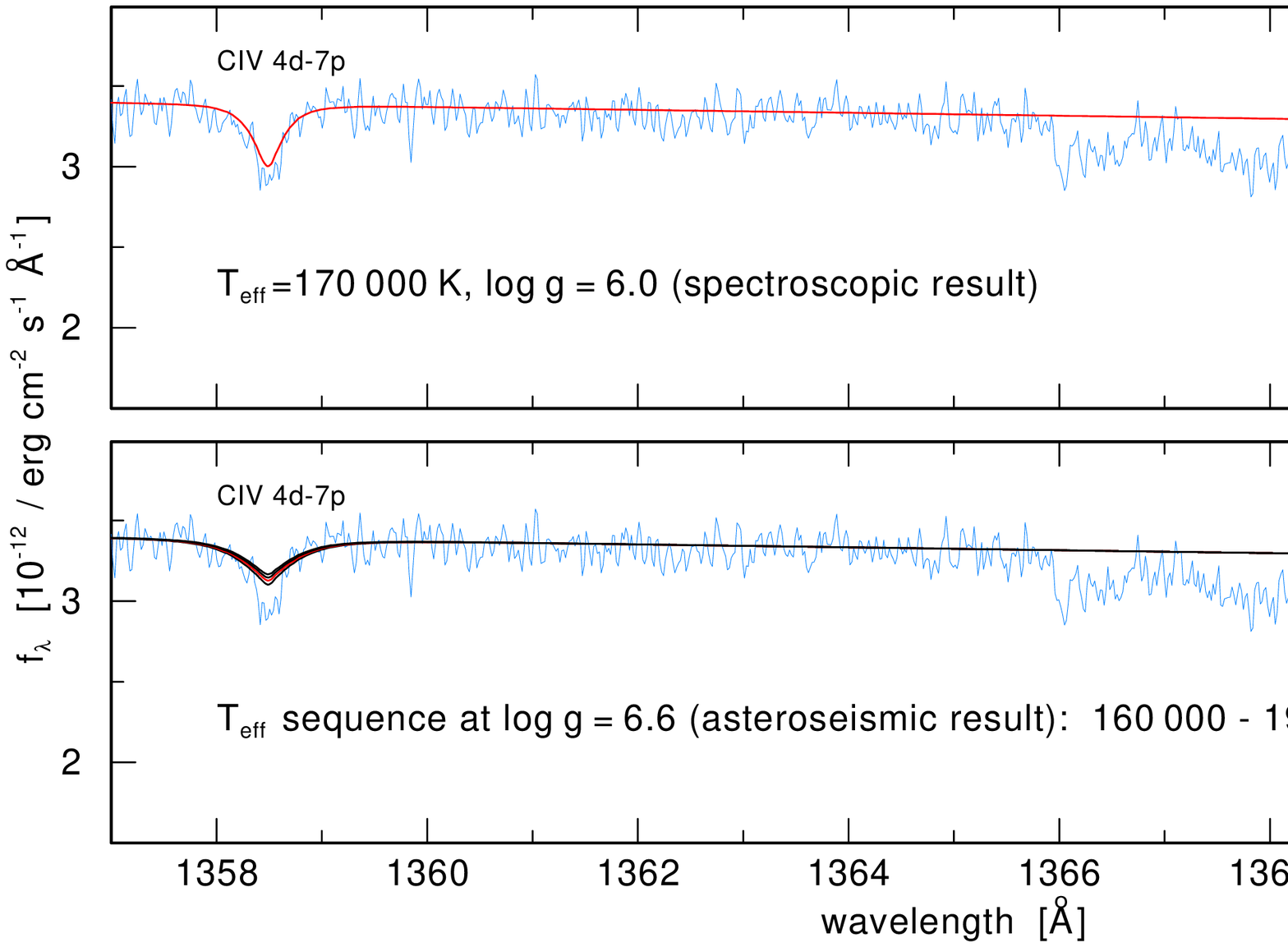}
\caption{Detail of a HST spectrum of \rxj\ comprising a 
\ion{C}{iv} line and the location
of the \Teff\ sensitive \ion{O}{v}~1371~\AA\ line. The \ion{O}{v} line
is not detected,  placing a lower limit to  \Teff. \emph{Upper panel:}
Comparison with  a synthetic spectrum  from a model  with \Teff=170~kK
and   \logg=6.0,  the   best-fit  parameters   from  optical   and  UV
spectroscopy.  At \Teff$\leq$170~kK the model predicts a detectable
\ion{O}{v} line. \emph{Lower panel:} Comparison with four synthetic
spectra with  \Teff=160, 170, 180,  190~kK.  The \ion{O}{v}  line gets
weaker with increasing \Teff.  The surface gravity is fixed at
\logg=6.6, the asteroseismic result.  At this higher gravity, \Teff\
must exceed  180~kK in order  to have non-detectable  \ion{O}{v} line,
significantly above the asteroseismic result of \Teff=163~kK.}
\label{fit}
\end{figure*}

The last column in Table  \ref{tabla1} shows the theoretical rate of 
period change
of the fitted pulsation  modes. Our calculations predict all pulsation
periods  to  \emph{decrease}  with  time  ($\dot{\Pi}_k<0$), an  effect
attributable to the rapid contraction experienced by the star in its
evolution to the  blue (see  Fig. \ref{teff-g}). Unfortunately, an  
unambiguous $\dot \Pi$ 
measurement for  any mode of
\rxj\ is currently  very difficult because of the large amplitude
variations of the  modes (VEA02).  On the basis of  considerations of the 
mass-loss rate and the helium-rich envelope mass in \rxj, VEA02 
give a rough  estimation of the rate of period change of 
$|\dot{\Pi}|  \sim 3  - 24  \times 10^{-11}$  s/s,  much in
agreement with our  predictions. 

\section{Discrepancy between spectroscopic and seismic mass}
\label{masses}

The main features of our best-fit model are summarized in Table
\ref{tabla2}\footnote{Errors in $T_{\rm eff}$ and 
$\log (L_*/ L_{\odot})$ are estimated from the width of the minimum 
in the function $\chi^2$ vs $T_{\rm eff}$ and $\log (L_*/ L_{\odot})$, 
respectively; the error in the stellar mass comes from the grid resolution 
in $M_*$. Errors in the other quantities are derived from 
these values.}, where we also include the parameters of \rxj\ 
from other  published studies. Note that the  effective 
temperature of
the   best-fit  model   is   compatible  at   $1   \sigma$  with   the
spectroscopically  inferred  value.   According  to our  results,  the
temperature  of \rxj\  is  somewhat   lower  than  previous
estimations, but  it is still  the hottest known PG1159  pulsator.  On
the other hand,  the best-fit model has a stellar  mass of $M_*= 0.565
M_{\odot}$, very  close to  the values found  from the  period spacing
data (see \S
\ref{period-spacing}).   \emph{Thus,  the  linear   pulsation  theory
strongly suggests  that \rxj\ is an average-mass  PG1159 star
with $M_*  \sim 0.56-0.57 M_{\odot}$}.   This is clearly at  odds with
the  spectroscopic-evolutionary predictions,  that  indicate a  rather
high mass of  $0.70 - 0.72 M_{\odot}$ for this  star (Werner \& Herwig
2006; Miller Bertolami  \& Althaus 2006).  A mass  discrepancy is also
encountered with  other PG1159 pulsators  (PG\,1159-035, PG\,1707+427,
PG\,2131+066, PG\,0122+200).   In those  cases the seismic  masses are
always \emph{higher}  than the spectroscopic masses  (see C\'orsico et
al.  2006), however,  the differences  are a  factor of  three smaller
($0.04-0.06 M_{\odot}$) than for \rxj.

The mass discrepancy for \rxj\ is reflected by the difference
between  the surface  gravity of  our  pulsation model  and the  value
predicted by  spectroscopy.  In fact,  for our best-fit model  we have
$\log g= 6.61$, large in excess  as compared with the derived value of
$\log g=  6.0^{+0.3}_{-0.2}$ by using NLTE  model atmospheres.  Hence,
the  location of  the star  in the  $\log T_{\rm  eff} -\log  g$ plane
shifts,  according to  our seismological  predictions,  downwards well
beyond the limits of the observational error box (see Fig. \ref{teff-g}).

It  is obvious  that  the seismic  results  on \Teff\  and \logg\  are
difficult to reconcile with results of model atmosphere analyses. This
is demonstrated with Fig.\,\ref{fit}, which shows a HST spectrum of
\rxj\ in the vicinity of the \ion{O}{v}~1371~\AA\ line. This feature 
is very  temperature sensitive  and is therefore  used as  a strategic
line to  fix \Teff\  in PG1159  stars. It is  not detectable  in \rxj\
which puts a tight lower limit to
\Teff.  The lower panel of Fig.\,\ref{fit} shows a series of model
profiles with increasing  \Teff, whereas the gravity is  kept fixed at
the  asteroseismic value  \logg= 6.6. The model atmospheres were 
calculated with the T\"ubingen Model Atmosphere Package 
(see, for details, Werner et al. 2003). The model  with \Teff=160~kK  (a
value close to  the asteroseismic result of 163~kK)  produces a strong
\ion{O}{v}  line,  clearly  contradicting  the  observation.  A  model
temperature  of 190~kK  is  necessary to  make  the line  undetectably
weak. In the upper panel  we show the model with the spectroscopically
determined  best-fit  parameters  (\Teff=170~kK, \logg=6.0).   In  the
analysis of  Werner et al. (1996)  we regarded this \Teff\  as a lower
limit but excluded temperatures higher than 180~kK because all
\ion{He}{ii} and \ion{C}{iv} lines in the optical and UV become too
shallow.   This discrepancy  between  spectroscopic and  asteroseismic
results calls for a  comprehensive re-analysis taking into account all
available spectroscopic material collected since the Werner et al. (1996) 
analysis.

VEA02 infer the stellar luminosity  and radius of \rxj\ on the
basis  of their  mass  estimation and  the  effective temperature  and
gravity  measured  by  spectroscopy.   Thus, not  surprisingly,  their
results point to a  larger luminosity and radius ($\log L_*/L_{\odot}=
4.05_{-0.32}^{+0.23}$             and            $\log(R_*/R_{\odot})=
-0.91_{-0.15}^{+0.10}$)  as compared  with our  best-fit  model ($\log
L_*/L_{\odot}= 3.36$ and $\log(R_*/R_{\odot})= -1.23$).

\section{Helium envelope thickness}
\label{helium}

\begin{figure}
\centering
\includegraphics[clip,width=250pt]{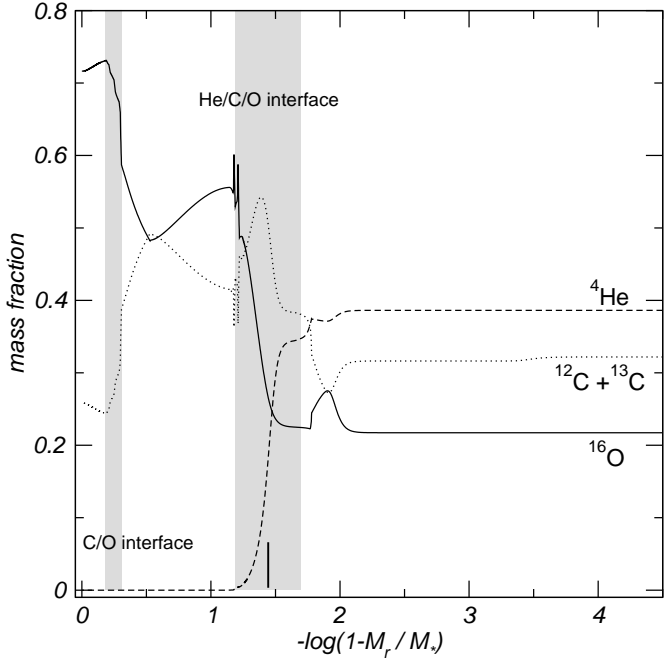}
\caption{The internal chemical profile of our best-fit model in terms 
of  the fractional mass  coordinate. The thickness of the outer envelope
is   $M_{\rm env}= 0.02  M_{\odot}$.  The location of the 
C/O and He/C/O chemical transition regions is emphasized 
with grey regions. The vertical line at $-\log(1-M_r/M_*)\sim 1.4$ 
marks the location of the layer where $X_{\rm He}\sim 0.19$.}
\label{perfil} 
\end{figure}

Another  important parameter  is  the thickness 
of the outer envelope ($M_{\rm  env}$)  of \rxj. 
We define $M_{\rm  env}$ as the mass above 
the chemical discontinuity between the He-rich envelope and the C/O 
core (see Fig. \ref{perfil}). Our  best-fit model has 
$M_{\rm env}= 0.02 M_{\odot}$\footnote{We define the 
``effective'' location of the He/C/O chemical interface as the position of the 
layer characterized by an helium abundance of $X_{\rm He} \sim 0.19$. 
In our best-fit model this corresponds at $M_r \sim 0.544 M_{\odot}$ 
(see the vertical line in Fig.  \ref{perfil}), 
where the contribution of the He/C/O chemical interface to the 
Brunt-V\"ais\"al\"a frequency is largest.}. 
VEA02 claim that $M_{\rm  env}$ is $0.0073-0.044M_{\odot}$,
which is in line with our results.  
It is important to note  that VEA02 derive
these values of $M_{\rm env}$ on  the basis of the results published by
Kawaler \& Bradley  (1994) for PG 1159-035, extrapolated  to the range
of parameters  of \rxj.   Specifically, they use a relation  
connecting the fractional  mass  of  the  helium-rich  envelope  with  
the  effective temperature   and  the  trapping   cycle  for   
PG\,1159-035   and  \rxj.  The  underlying  assumption  is  
that  the  mode-trapping features of  PG1159 stars are  inflicted 
mainly by the  outer chemical
interface of He/C/O,  and thus, the minima seen  in the period-spacing
distribution are  associated with modes trapped in  the envelopes. 
However, realistic evolutionary calculations predict mode-trapping
properties of PG1159  models to be primarily  fixed by  the 
stepped structure  of the  C/O  chemical profile  at  the core  left by  
prior extra mixing episodes (Fig. \ref{perfil}),   the  outer  
He/C/O   interface  playing  a secondary role  (see C\'orsico \&  
Althaus 2005, 2006).  For instance,
the observed  periods at 799.50 s  ($k= 35$) and 1058.03  s ($k= 47$),
which correspond  to modes  trapped in the  envelope of  \rxj\
according  VEA02, are  modes ``confined''  to the  core region  in our
best-fit model  and thus not related  to the thickness of the helium-rich  
outer  envelope.   Thus,  the   procedure from which VEA02  derived 
the thickness of the helium-rich outer envelope of \rxj\ should be
taken with some caution.  

\begin{table*}
\centering
\caption{The main characteristics of the central star and 
the planetary nebula of \rxj. The second column  
corresponds to spectroscopic results, whereas the third and fourth 
columns present results from the pulsation study of VEA02 and 
from the asteroseismological model of this work, respectively.} 
\begin{tabular}{l|ccc}
\hline
\hline
Quantity & Spectroscopy & Pulsations & Asteroseismology \\
         &              &    (VEA02) & (This work)      \\ 
\hline
$\overline{\Delta \Pi}$ [s] &  --- & $21.618\pm 0.008$     &  21.707  \\ 
\hline
$T_{\rm eff}$ [kK] & $170 \pm 10^{\rm (a)}$                &        ---                   & $163.4_{-3.7}^{+2.5}$ \\
$M_*$ [$M_{\odot}$] & $0.72_{-0.12}^{+0.15\rm (b)}$        &  $0.56_{-0.04}^{+0.02}$      & $0.565_{-0.023}^{+0.024}$ \\ 
$\log g$ [cm/s$^2$] & $6.0_{-0.2}^{+0.3\rm (a)} $          &        ---                   & $6.61_{-0.07}^{+0.11}$  \\ 
$\log (L_*/L_{\odot})$ & $3.95 \pm 0.5^{\rm (c)}$          &  $4.05_{-0.32}^{+0.23}$      & $3.36 \pm 0.04$ \\  
$\log(R_*/R_{\odot})$ &  ---                               &  $-0.91_{-0.15}^{+0.10}$     & $-1.23_{-0.025}^{+0.046} $ \\  
$M_{\rm env}$ [$M_{\odot}$]  & ---                         &  $0.0073-0.044$              & $0.02 \pm 0.006$ \\  
$X_{\rm He},X_{\rm C},X_{\rm O}$ &   0.39,  0.55,  0.06    &        ---                   & 0.39, 0.32,  0.22  \\   
BC [mag]                       & ---                     &        ---                     & $-7.954_{-0.16}^{+0.01}$\\   
$M_{\rm V}$ [mag]                & ---                     &        ---                   & $4.43_{-0.23}^{+0.12}$\\
$A_{\rm V}$ [mag]                & ---                     &  0.86                        & $0.45_{-0.012}^{+0.036}$\\
$d$  [pc]  & $1400^{+700{\rm  (c)}}_{-500}$                &  $760_{-235}^{+230}$         & $452_{-23}^{+46}$  \\ 
$\pi$ [mas]&  $0.71_{-0.24}^{+0.40}$                       &  $1.32_{-0.30}^{+0.59}$      & $2.21_{-0.20}^{+0.12}$\\ 
$t_*$ [yr] & $2 - 5 \times 10^{4{\rm  (d)}}$               &  $> 1.3 \times 10^{5}$       & $\gtrsim 2.5 \times 10^4$\\
\hline
$D_{\rm PN}$ [pc] & $5.3_{-1.9}^{+2.7\rm  (d)}$                       &  $2.9 \pm 0.9$               & $1.72_{-0.09}^{+0.18}$\\
$t_{\rm PN}$ [yr] & $\gtrsim 1.5 \times 10^{5{\rm  (d)}}$  &  $\sim 9.1 \times 10^4$(*)   & $5.43_{-0.29}^{+0.57} \times 10^4$\\
\hline
\hline
\end{tabular}
\label{tabla2}

{\footnotesize  References: (a)  Werner et al. (1996);  (b) Miller
Bertolami \& Althaus (2006); (c) Motch et al. (1993); 
(d) Appleton et al. (1993). (*) This value is inferred from 
the linear size of the planetary nebula derived by VEA02 
and a mean expansion velocity of the nebula of 31 km/s.}
\end{table*}

The employment of full PG1159 evolutionary models allows us to place
constraints on the amount of stellar mass eroded by winds. We can do
this by adopting typical mass loss rates for the different stages
after the VLTP. For the short superwind phase, that lasts $\sim800$ yr
and takes place at \lteff $<3.8$, we adopt $10^{-5}$ $M_{\odot}$ /yr as
observed at Sakurai's Object (Tyne et al. 2002). For the evolution
before the star reaches \Teff$\sim 100\, 000$ K (this stage takes
$\sim$2000 yr) we adopt typical [WR] planetary nebula nuclei mass loss
rates of $10^{-6}$ $M_{\odot}$/yr (Koesterke 2001). Finally, for
the PG1159 stage we can adopt the observed value of $4 \times 10^{-8}$
$M_{\odot}$/yr (during the 12000 yr until the sequence reaches the
best fit model). Under these assumptions we estimate the possible mass
lost by the star after the VLTP to be $10^{-2}$ $M_{\odot}$, which is
half of the mass of the He-rich envelope. Thus the remaining envelope
will be still thick enough for it not to affect the results of the
present work. Even more, as the evolution of the star would be
accelerated by these mass loss episodes, the previous estimation represents
an upper limit under the adopted mass loss rates.

\section{The seismic distance to \rxj\ and the 
implications for its planetary nebula}
\label{pn}

In closing, we infer the seismic distance of \rxj\ from the
Earth. First, we consider the flux predicted by a NLTE model atmosphere 
with $T_{\rm eff}= 160$ kK and $\log g= 6.6$ integrated through the 
spectral response of the V filter. We obtain a bolometric correction 
BC= $-7.954$ and an absolute magnitude  
$M_{\rm v}= 4.43$. Because the proximity of \rxj\ 
to the galactic plane, we must account for the interstellar absorption, 
$A_{\rm V}$. In 
line with VEA02, we follow the interstellar extinction model developed 
by Chen et al. (1998). We compute the seismic distance $d$ according 
to the well-known relation: 

\begin{equation} 
\log d = \frac{1}{5} \left[ m_{\rm v} - 
M_{\rm v} +5 - A_{\rm V}(d) \right], 
\label{distance}
\end{equation}

\noindent where the apparent magnitude is $m_{\rm v}= 13.16 \pm 0.01$ 
(Motch et al.  1993). The interstellar  absorption $A_{\rm V}(d)$ varies 
non linearly with the distance and also depends on the Galactic 
latitude ($b$) and 
longitude ($\ell$). For the equatorial coordinates of \rxj\ 
(Epoch B2000.00, $\alpha=  21^{\rm h}\ 17^{\rm m}\ 7^{\rm s}.60$, 
$\delta= +34^{\circ}\ 12'\ 22''.0$)
the corresponding Galactic coordinates are $b= -10^{\circ}\ 24'\ 32''.04$ 
and $\ell= 80^{\circ}\ 21'\ 10''.8$. We solve Eq. (\ref{distance}) 
iteratively and obtain a 
distance  $d= 452$  pc and an interstellar extinction  
$A_{\rm V}= 0.45$, substantially lower than the estimations of 
VEA02  ($d= 760^{+230}_{-235}$ pc and $A_{\rm V}= 0.86$).   
This is due  to the fact  that the
luminosity of our model  ($2.3 \times 10^3 L_{\odot}$) is considerably
lower than the  luminosity adopted by those authors  ($1.1 \times 10^4
L_{\odot}$). 

Our calculations predict a parallax of $\sim 2.21$ mas and a linear
diameter of the planetary nebula of $\sim 1.72$ pc ($5.3 \times
10^{13}$ km). If we assume a mean expansion velocity for a typical
planetary nebula with a WR-type central star of approximately 31 km/s
(Gorny \& Stasinska 1995), then the expansion age for the nebula would
be $t_{\rm PN} \approx 5.43 \times 10^4$ yr. On the other hand, our
computations predict an evolutionary age for the best-fit VLTP model
of about $t_* \approx 2.5 \times 10^4$ yr and possible larger if the
VLTP would have occurred later on the hot white dwarf cooling branch than 
in our sequence\footnote{We mention that our $0.565 M_{\odot}$ sequence
predicts a time interval between the occurrence of the VLTP and the
present state of \rxj\ of $\approx 1.5 \times 10^4$ yr.}.  This age is
well consistent (to a factor less than two) with the expansion age of
the nebula. Thus, our results seem to solve the apparent inconsistency
between the evolutionary timescale of \rxj\ and the size of the nebula
(see Appleton et al. 1993 for details).  The agreement between $t_*$
and $t_{\rm PN}$ according to the present calculations reinforces the
validity of our asteroseismological model and may suggest that \rxj\
have undergone a VLTP episode, a fact that would be at variance with
the lack of $^{14}$N in its atmosphere.

\section{Summary and conclusions}
\label{conclusions}

In this  paper we carried  out an asteroseismological analysis  of the
hot  pulsating PG1159  star \rxj, a $g$-mode pulsator with 
properties intermediate between young planetary nebula central stars
and PG1159 stars without planetary nebula. Our analysis is 
based on  the full  PG1159 evolutionary  models of  
Miller  Bertolami \&  Althaus (2006).   These models  represent  a  
solid  basis  to analyze  the  evolutionary  and
pulsational  status  of  hot   PG1159  stars  like \rxj,  a
transition object  for which stellar models  extracted from artificial
evolutionary procedures  are not appropriate.  We first  made good use
of the fact that the period spacing of variable PG1159 stars is mainly
a  function of  the  stellar mass,  and  derived a  value of  $M_*\sim
0.568M_{\odot}$ by comparing $\Delta  \Pi^{\rm O}$ with the asymptotic
period spacing of  our models.  We also compared  $\Delta \Pi^{\rm O}$
with  the  computed period  spacing  averaged  over  the period  range
observed  in \rxj,  and   derived  a  value   of  $M_*\sim
0.560M_{\odot}$. Note that in both  derivations of the stellar mass we
made  use  of  the spectroscopic constraint  that  the
effective  temperature of the  star should  be of  $\sim 170$ kK. 

Next, we adopted a less  conservative approach in which the individual
pulsation periods  alone naturally lead  to an ``asteroseismological''
PG1159 model  that is assumed  to be representative of \rxj.
The period fit was  made on a grid of PG1159 models  with a quite fine
resolution in effective temperature ($\Delta T_{\rm eff}\sim 10-30$ K)
although admittedly coarse  in stellar mass ($\Delta M_*  \sim 0.012 -
0.077  M_{\odot}$). The  match between  the computed  dipole pulsation
periods  of  the  best-fit  model  and  the  observed  periods  in  
\rxj\  is  excellent, with  an  average  of  the absolute  period
differences of $1.08$  s and a root-mean-square residual  of $1.34$ s.
It is  worth noting also that  the domain of  unstable-mode periods of
the best-fit model nearly matches the range of the observed periods in
\rxj,  although there  is no obvious correlation  between the
magnitude of the linear growth rates and the observed mode amplitudes.

Interestingly  enough, the  mass of  the best-fit  model  ($M_*= 0.565
M_{\odot}$)  agrees with  the mass  derived from  the  observed period
spacing,  but it is $\sim  25 \%$  lower than  the value  $M_*= 0.72
M_{\odot}$  derived from  spectroscopy coupled  with  our evolutionary
tracks.

Other characteristics of the best-fit model are summarized in Table
\ref{tabla2}. In particular, the effective temperature is compatible
at $1  \sigma$ with the  spectroscopic value.  At variance  with this,
the surface gravity of the best-fit model is substantially higher than
that given by spectroscopy.  We also infer the ``seismic distance'' of
\rxj\ and obtain
a distance  $d \sim 452$ pc,  which places the star  markedly closer to
the Earth than thought hitherto  ($d\sim 760-1400$ pc, Vauclair et al.
2002 and Motch  et al.  1993).  The derived  distance implies a linear
size of the nebula of 1.72 pc which implies an expansion age
of  $\approx 5.43  \times 10^4$  yr. This age is substantially lower 
than assumed hitherto (Appleton et al. 1993) and in better 
agreement with  the times predicted by evolutionary models. 
This result reinforces the correctness of our asteroseismological 
model for \rxj.

Finally,  our computations  predict a
temporal period  drift for \rxj\ between  $-3\times 10^{-11}$
s/s and $-1.4 \times 10^{-10}$ s/s. The negative values of $\dot{\Pi}$
reflect the fact that our  best-fit model is still rapidly contracting
on its evolutionary road  towards  higher effective temperatures in
the  HR  diagram.  Unfortunately,  the amplitude  variability  of  the
observed modes in \rxj\  precludes any measurement of $\dot{\Pi}$
for the moment, thus hindering any test of our prediction.

The results  of the asteroseismological  analysis carried out  in this
work strongly suggest  that the mass of \rxj\ is considerably
lower  than  suggested by  spectroscopy  coupled  to   evolutionary
tracks (Werner \& Herwig 2006; Miller Bertolami \& Althaus 2006).  
This serious  disagreement is  also seen  in  other pulsating
PG1159  stars  like  PG\,1159-035,  PG\,1707+427,  PG\,2131+066  and  
PG\,0122+200, although in those cases the seismic masses are always higher
than  the spectroscopic  masses (C\'orsico et al. 2006).  The  discrepancy 
could  be  attributed  to  a number  of  factors.  On  the
observational side, possible errors in the spectroscopic determinations
of  $g$ and $T_{\rm  eff}$; in particular, as this study suggests in the
case of \rxj, in the determination of $g$. In this respect, detailed 
asteroseismological fits to other pulsating PG1159 stars would be valuable.
On  the   theoretical  front, different PG1159 evolutionary tracks 
could result from different dredge-up and/or mass loss history in the 
AGB progenitor evolution (Werner \& Herwig 2006). But in preliminary 
simulations we have found that neither
third dredge up efficiency nor TP-AGB lifetimes play a determining role 
in the location of PG1159 tracks. It remains to
be seen if other assumptions in the microphysics such as radiative and 
conductive opacities and/or equation of state may be playing a role 
in the location of post-AGB tracks. 


\begin{acknowledgements}
We  warmly acknowledge our  referee (S. O. Kepler) for  his
comments  and suggestions which strongly improved the original version  
of the paper. We thank Rub\'en Mart\'{\i}nez and H\'ector Viturro for 
valuable technical support. M.M.M.B. thanks an EARA-EST from the European 
Association for Research in Astronomy. This research was partially supported 
by the PIP 6521 grant from CONICET.
\end{acknowledgements}

\end{document}